\documentstyle[aps,epsfig]{revtex}
\begin{document}

\title{Neutron Catalysis of Resonance Fusion in Stellar Matter}

\author{Nurgali Takibayev \thanks{\textit{E-mail address:}
teta@nursat.kz}}

\address{Department of Physics \& Mathematics, \\
Kazakh National Pedagogic University, Almaty, Kazakhstan}

\maketitle


\begin{abstract}

Within the framework of resonance fusion study
in stellar matter the features of $n,\alpha,\alpha$ - system  have
been investigated at astrophysical energies.
Consideration of three body scattering has been carried out on base
of well-known Faddeev's equations.

It is found that under certain conditions the series of resonance
states appear in $n,\alpha,\alpha$ system at very low energies
($\sim keV$ ). The lifetimes of these three body resonances are
close to the lifetime of unstable nucleus $^8Be$.

The ground state of $\alpha,\alpha$ - system is regarded as very
narrow resonance with the energy $\sim 92$ keV and the width $\sim
6.8$ eV. The simple forms of two body repulsive potentials are
taken into account to describe the parameters of the
$\alpha,\alpha$ - resonance and to satisfy $n,\alpha$ scattering
data at very low energies.

The explanation of resonance phenomena in $n,\alpha,\alpha$ -
system is offered on base of physical model. The effect
results from resonance quantum phenomena in few body dynamics.

In turn, the resonance
fusion can give influence on many astrophysical phenomena.
The possibility of catalyzing this new mode of fusion
by free neutrons in $\alpha$-particle matter is considered too in present
exposal.
\end{abstract}


\section*{Introduction}

It is the well-known fact that F. Hoyle predicted the existence of
narrow low energy resonance in the system of three $\alpha$-particles
\cite{Hoyle}. Then his prediction was confirmed by experiments. It
caused the essential progress of nucleogenesis understanding in
stellar interior  \cite{Fowler} -  \cite{Streigman}.

The present exposal contributes to development of low energy resonance
astrophysics, in particular, resonance phenomena in system consisted of
two $\alpha$ - particles and one neutron. Consideration has been
given to the new type of fusion reactions, which can take place in
stellar matter. These reactions are resulted from resonance quantum
phenomena in few body systems  \cite{FB19} -
\cite{PenkovTakibayev}.

Note that neutron and $\alpha$ - particle can not form a
bound state. Neutron of very low energy can not be captured by $\alpha$
- particle. Its effective potential of interaction is
characterized as a repulsing power in S-wave component. However,
attraction in P-wave components is sufficiently strong to
result in resonance behavior of corresponding amplitudes
near the energy of $\sim 1 \ MeV$  \cite{Ajzenberg-Selone90}.

Two $\alpha$ - particles can not be bound in a stable state but in
scattering they have the very narrow resonance at low energy
region. The resonance energy $\simeq 92 \ keV$ and the width $\simeq 6.8
 \ eV$ \cite{Ajzenberg-Selone90}. Moreover, the system of three
$\alpha$-particles has
the resonance state of long lifetime too. The resonance energy and
width are $\simeq 0.38 \ MeV$ and $\simeq 8.5 \ eV$, respectively, in
this case  \cite{Ajzenberg-Selone506}.

We have found that three body system consisting of two
$\alpha$-particles and one neutron has a set of narrow resonances
situated under the ground state energy of their subsystem - $^8Be$
\cite{FB19}. The last one is considered as a sharp resonance.

In the system of three $\alpha$-particles and one neutron the number of
similar resonances may be even more.
So, it is proposed that a free neutron can gather few
$\alpha$-particles to form quasi-molecular resonance states.

As for systems of few $\alpha$-particles the probability of
resonance state formation at very low energies is small,
and Coulomb barrier is also preventing it.
Accompaniment of free neutron to the system of few $\alpha$ - particles
mainly changes a situation. Amongst $\alpha$ - particles a neutron
will play a role of interchanging particle, creating efficient
additional attraction.

Therefore owing to free neutron the low energy fusions become possible
with few $\alpha$-particles. They can product the different
nuclei including nuclei with mass several times superior
than $\alpha$ - particle's mass.

The neutron can get free and then stimulate resonance reactions
again. In the fusion reactions the neutron can be released taking a
lot of the kinetic energy. The neutron is slowed down in an
ambience of $\alpha$ - particles and then can form once again
resonance quasi-molecular states.

Mechanism of neutron catalysis in resonance fusion reminds well
known $\mu$ - catalysis in nuclear fusion of $\mu$, d, t -
molecules  \cite{Gernshtein}.

It is assumed that inside stars there is presence of area with the
high density of helium, where can occur described processes.

The resonance fusion can have influence on many astrophysical
phenomena. Some of them are discussed below.

\section{General restrictions and Two body potentials}

We consider a very low energy region - close to the temperature at
the center of the Sun ($ \ll 1 MeV$). Relativistic effects, contributions of
inelastic nuclear channels, quark structure of nucleons and nuclei
etc are not taken into account. Nucleons and $\alpha$-particles are
considered as nonstructural bodies. Effects of rearrangement
between neutrons and nuclei are ignored. Thus, it is under
consideration a very simple model to investigate the resonant
phenomenon in three body dynamics.

In case of simple isolated resonance in two body system the
amplitude of scattering has the distinctive form
\begin{equation}
\label{fres} f(E)\simeq
\frac{-1}{\pi\rho(E)}\frac{\Gamma/2}{E-E_R+i\Gamma/2}.
\end{equation}
It means that two-body amplitude has poles in $k$-complex plane in points of
$k=k_{Res}=\pm k_R+ik_I$, where $E_{Res}=k_{Res}^2/2\mu$,
$E_R=(k_R^2-k_I^2)/2\mu$, $\Gamma=-4k_Rk_I/2\mu$  and $k_I < 0$,
$\mu$ - reduced mass, $\rho(E)$ - the density of states in the
continuum region.

Interactions between two particles are chosen in the simplest form
- in form of separable potentials:
\begin{equation}
\label{VYam} \{\lambda V \}_{L,S}^J=|\nu_{L,S}^J> \lambda_{L,S}^J
<\nu_{L,S}^J|.
\end{equation}
In this expression only a partial component of potential is written out. They
satisfactory describe resonance behavior in pair subsystems
\cite{Blokhintsev}. Hereinafter the angular functions, recoupling
coefficients, etc are omitted for the simplification. The quantum
scattering theory of three-body system is given an account of
\cite{Faddeev}. Theory of quantum few-body systems with details can
be found, for instance, in  \cite{Glockle}.

Using the separable potential gives an exact analytical solution of
two-body problem. It is their obvious quality. The scattering
amplitude can be written as:
\begin{equation}
\label{fYam} f(k,k';Z)= <k |\hat{f}|k' >, \  \
 \hat{f}=-|\nu> \eta(Z) <\nu|, \\
\end{equation}
where
\begin{equation}
\label{AYam}
 \eta^{-1}=\lambda^{-1}-A(Z), \  \  A=<\nu|G_0 (Z)|\nu>,
\end{equation}
and for simplicity indexes of states are omitted too. The amplitude
$f(k,k';Z)$ comes to the physical one $f(E)$ in limit $Z\rightarrow
E=k_0^2/2\mu$, where $E$ - the real quantity
 and $|\vec{k}|=|\vec{k'}|\rightarrow k_0$.

In case of potential (\ref{VYam}) is represented as a sum of separable ones
the expressions of (\ref{fYam}) and (\ref{AYam}) have to be understood as
matrix values in relation to corresponding index.

The partial amplitude can get a pole in point of bound state
$Z=E_B$, when $\lambda^{-1}=A(E_B)$ and $E_B \leq 0$. But a pole
may be in point of resonance state $Z=E_{Res}$, when
$\lambda^{-1}=A(E_{Res})$  and $E_{Res}$  is  a complex quantity
 \cite{TakibayevDubna}.

It is convenient to normalize form-factors with condition $A(0)= - 1$
appointing $\lambda$ as dimensionless constant.

\subsection{The separable $n,\alpha$-potentials}

Effective potential of $n,\alpha$-interaction at low energy region
has the following properties. Its S-wave component corresponds to
repulsive force in accordance with Pauli principle. The S-wave
potential represents the main factor in our case,
i.e. at very low energies.

For S wave the form-factor of potential (\ref{VYam}) can be chosen as:
\begin{equation}
\label{nSYam} <\nu_{n,\alpha}|k> = N^S_{n,\alpha}/(1+x^2),
\end{equation}
where the constant $N^S_{n,\alpha} = \sqrt{8\pi/(2\mu\beta)}$, the variable
$x=k/\beta$ and $\beta$ -
the inverse radius of nuclear force.

It should be noted that in case of attractive force the
equation $\lambda^{-1}=A(E_B)$ gives a bound energy  $E_B = -
k_I^2/2\mu$, where $x_I = k_I/\beta = \kappa$, $\lambda = - (1 +
\kappa)^2$,  $\kappa > 0$ and $\lambda \leq - 1$ .

In case of $- 1 < \lambda < 0$ one can obtain a quantity for virtual
state: $\kappa = - 1 + \sqrt{-\lambda} < 0$.

In the contrary in case of the repulsive potential (\ref{VYam})
with form-factor (\ref{nSYam}) the resonance
state appears in the system. Potential parameters obey the following
relationship: $x_R^2 = k_R^2/\beta^2 = \lambda > 0$ and $k_I =
-\beta < 0$, which results from equation:
$\lambda^{-1}=A(E_{Res})$.

So, the parameters fixed by $n,\alpha$
experimental scattering data are equal: $\beta = 0.751 fm^{-1}$ and
$\lambda = 14.94$. In turn they give the resonant energy and width
such as: $E_R \sim \Gamma \sim 200 \ MeV$ \cite{TakibayevDubna},
\cite{Blokhintsev}.

Other components of effective $n,\alpha$-interaction correspond to
the attraction forces, and the most important among them are P-wave
interactions.

For P-waves ($P_{L,S}^J$, with $L=1$, $S=1/2$ and $J=3/2$ or
$J=1/2$) the form-factor may be chosen, for example, in following forms:
\begin{equation}
\label{nPYam} <\nu^P_{n,\alpha}|k> = N^P_{n,\alpha} \frac{x}{1+x^2} \ , \  \  \  or
\  \ <\tilde{\nu}^P_{n,\alpha}|k> = \tilde{N}^P_{n,\alpha} \frac{x}{(1+x^2)^2}  \ .
\end{equation}

Then, parameters of $n,\alpha$-system have to be in agreement with
experimental data. For first of them ($J = 3/2$) they are following:
$x_R = k_R/\beta \simeq 0.158$, $x_I = k_I/\beta \simeq - 0.0256$,
$\beta = 1.175 fm^{-1}$ and $\lambda = - 0.97$ that correspond to
$E_R \simeq 0.9 MeV$ and $\Gamma \simeq 0.6 MeV$.
 It is easy to
determine all parameters for whatever other case and another selection.

The description of low energy $n,\alpha$-scattering is sufficiently
assigned by potentials of  S- and P-waves up to energies $\simeq 10
\ MeV$ \cite{Blokhintsev90},  \cite{Arndt}.

\subsection{The separable $\alpha,\alpha$-potentials}

As it is known, no stable nuclei are consisted of 5 or 8
nucleons. For example, $^8Be$ has a lifetime $\sim 10^{-16} s$
\cite{Ajzenberg-Selone90}.
Obviously, it is a huge time in relation to nuclear time scale.

Experimental data give
parameters of this resonance: parity  $0^+$, energy  $92 \ keV$ and
width $6.8 \ eV$ \cite{Ajzenberg-Selone90}.
Note that the experimental data of $\alpha,\alpha$-resonance give
the ratio
\begin{equation}
\label{kIR} |k_I/k_R| \simeq 0.185\cdot10^{-4} \ .
\end{equation}

In general, effective potential of two $\alpha$-particles has to be
a result of several complex interactions and involved forces. It
is clear that a determination of exact potential is a very
difficult problem of few-body physics to get solution even at low
energies.

However, it is
possible to describe this resonance within the framework of simple
potential model like (\ref{VYam}).

We consider this nucleus as a very narrow resonance
in amplitude of
two $\alpha$-particle $S$-wave scattering at low energies.
That is why we will take into account only the corresponding part
of total two
body interaction and ignore other parts at very low energy region.
These parts can become essential at higher energies. We'll mark here the
resonance state as $(\alpha,\alpha)^{*}$.

Below we can see that parameter of resonant quality (\ref{kIR}) brings
about suppression of such interactions.

Moreover, when viewed of interaction between a neutron and the
$(\alpha,\alpha)^*$-system we ignore the Coulomb force too. We guess
that resonant phenomena in $n,\alpha,\alpha$-system depend on
features of neutron as neutral particle and
$(\alpha,\alpha)^*$-subsystem as resonator.

Indeed, Coulomb repulsion is of particular importance for
low-energy resonances, since it reduces the probability for charged
particles to approach each other closely and, hence, the
probability of their nuclear interactions. Moreover, Coulomb
repulsion can deform a resonance substantially - for example, push
it to the region of higher energies
 \cite{Baz}, \cite{Takib2005}.

Note, that problems of scattering governed by both Coulomb and
nuclear forces have special features inherent in them that are due
to the "bad" behavior of Coulomb forces in the asymptotic region.
This leads to the interplay of the nuclear and Coulomb transition
operators, with the result that, even upon the separation of the
purely Coulomb component from the total two-particle amplitude, the
remainder involves irregular factors.

The situation becomes even
more complicated in the problems where the scattering process
involves three or more particles (see, for instance
\cite{FaddeevMerkur}, \cite{Watson}).

However, the problem of describing the properties of resonance
scattering becomes more tractable in the energy region around a
narrow resonance. First of all, this applies to a determination of
the Coulomb shifts of resonance levels  \cite{TakibayevDubna}.

Our calculations carried out with Coulomb force between two
$\alpha$-particles give a result of very small shift of resonance
level in case of repulsive shot-range potentials \cite{TakibAGU}.

Meanwhile in case of
attractive shot-range potentials the Coulomb shifts of resonance
levels can be very essensial \cite{Takib2005}.

In any case the resonant quality (\ref{kIR}) can not be worse without Coulomb
repulsive forces. It is a reason to consider
a simple model of repulsive shot-range potentials ignoring Coulomb forces
at first step of investigations.

Using S-wave potential form-factors for $(\alpha,\alpha)^*$-system in form:
\begin{equation}
\label{aSYam} <\nu^{(m)}_{\alpha,\alpha}|k> = N^{(m)}_{\alpha,\alpha}/(1+x^2)^{m/2} \ ,
\end{equation}
where $m = 2, 3, 4, 5$ and $6$, we can get exact solutions (\ref{fYam}) and then
determine corresponding potential parameters and quantities $x_R$, $x_I$ are
being in accordance with experimental data.

As far as  $m = 2$ the form-factor (\ref{aSYam}) gives unreasonable values of
$\beta$ and $\lambda$. In case of $m \geq 4$ parameters are more
reasonable and within the sensible limits (see Table 1).

\vspace{0.5cm}

\begin{table}
\begin{center}

\caption{Parameters of $(\alpha,\alpha)^*$ potentials with form-factors (8)}
{and corresponding resonance parameters.}
\end{center}
\begin{center}

\begin{tabular}{|c|c|c|c|c|c|}
m            & 2 & 3 & 4 & 5 & 6 \\
\hline \\
$\lambda$ & $2.9\cdot10^9$ & $1.08\cdot10^4$ & $9.06\cdot10^2$ & $3.13\cdot10^2$ & $1.75\cdot10^2$ \\
\hline \\
$\beta_{\alpha,\alpha} (fm^{-1})$ & $1.75\cdot10^{-6}$ & $1.58\cdot10^{-3}$ & $0.7\cdot10^{-2}$ & $1.4\cdot10^{-2}$ & $2.12\cdot10^{-2}$ \\
\hline \\
$x_R$ & $5.4\cdot10^4$ & 60.0 & 13.5 & 6.73 & 4.46 \\
\hline \\
$- x_I$ & 1 & $1.11\cdot10^{-3}$ & $2.5\cdot10^{-4}$ & $1.25\cdot10^{-4}$ & $0.827\cdot10^{-4}$ \\
\end{tabular}
\end{center}
\end{table}

\nopagebreak

\section{Three interacting bodies}

The quantum theory of three body scattering is excellently presented
in classical monographs and reviews on few-bodies and related subjects
(see, for instance  \cite{Faddeev},  \cite{Glockle},
 \cite{Baz},  \cite{FaddeevMerkur}, \cite{Belyaev} - \cite{Sandhas}).

 Here it's enough to
refer to needed formulas only for understanding and using in
calculations.

So, the three-body system can be considered as the system of particles
with simple pair interactions in form (\ref{VYam}) with parameters specified
in the preceding section. As usual, consideration has been carried
out within the framework of Faddeev's equations but using more
convenient way - approach of effective potential  \cite{Sandhas},
 \cite{Yadernaya Fizika}.

The amplitude of scattering of neutron on $\alpha,\alpha$-subsystem
can be written as:
\begin{equation}
\label{fij} f_{i,j} = - <\varphi^-_{\alpha,\alpha'}|(V_{n,\alpha} +
V_{n,\alpha'}) |\Psi^+_j> \ ,
\end{equation}
where $\varphi$ corresponds to the wave-function with one
two-body interaction only, $\Psi^{\pm}_j$ - a total wave-function, where
indexes
 $+(-)$ correspond to outcoming (incoming) waves. A total potential
in our case is a sum of two-body potentials: $V = V_{\alpha,\alpha'} +
V_{n,\alpha} + V_{n,\alpha'} = \sum V_i$, $ i=1,2,3 $. For example,
index $j$ in (\ref{fij}) corresponds to  final subsystem, i.e. marked a
number of first particle left the interaction field in asimptotic.

Introducing $Q_{i,j} = V_i|\Psi_j>$ we get Faddeev's equations in
form:

\begin{equation}
\label{Qij} Q_{i,j} = \tau_i \delta_{i,j} + \sum_{l=1}^{3}t_i G_0
\overline{\delta}_{i,l} Q_{l,j}  \ ,
\end{equation}
where $\tau_i = V_i|\varphi_i>$, $t_i$ - the pair
transition operator obeys $t_i = V_i + V_i G_0 t_i$,
the symbol
 $\overline{\delta}_{i,j}= I - \delta_{i,j}$ \
($\overline{\delta}_{i,j}=0$ if $i = j$ but
$\overline{\delta}_{i,j}=1$ if $i \neq j$).

Using a separable form of pair-interactions and introducing P-matrix
with expression
\begin{equation}
\label{QPij} Q_{i,j} - \tau_i \delta_{i,j} = |\nu_i> \eta_i P_{i,j}
\lambda_j <\nu_j|\varphi_j>
\end{equation}
one can end up to simple equations
\begin{equation}
\label{Pij} P_{i,j} = \Lambda_{i,j} + \sum_{l=1}^{3} \Lambda_{i,l}
\eta_l P_{l,j} \ ,
\end{equation}
where
\begin{equation}
\label{Lij} \Lambda_{i,j} = <\nu_i| G_0 (Z) |\nu_j> \ , \ \ with \ \ \
i \neq j .
\end{equation}
The sum on intermeduate states is intended, of course. For example,
equations (\ref{Pij}) can be written as:
\begin{equation}
\label{Pijtot} P_{i,j}(\vec{k}_i,\vec{k}_j) = \Lambda_{i,j}(\vec{k}_i,
\vec{k}_j) + \sum_{l=1}^{3}\sum_{\vec{k}_l}
\Lambda_{i,l}(\vec{k}_i, \vec{k}_l) \eta_l(Z_l) P_{l,j}(\vec{k}_l,
\vec{k}_j)
\end{equation}
Here $Z$ corresponds to the total energy of three body system, $Z_l = Z -
k_l^2/2\mu_l$ - subsystem energy of intermediate $\varphi_l$ state
and
\begin{equation}
\label{AlYam}
 \eta_l^{-1}=\lambda_l^{-1}-A_l(Z_l), \  \  A_l = <\nu_l|G_0 (Z_l)|\nu_l>.
\end{equation}

Equations (\ref{Pij}) and (\ref{Pijtot}) for transition matrix have to be understood
now as a set of equations for transitions between different
two-body channels. In that the $\alpha,\alpha$-resonance state is
meant as the excited body. In this case it has to be taken into
account rearrangement of particles, factors of angular, spin and
other quantities for two- and three-body states, recoupling
coefficients in momentum space, etc  \cite{Belyaev}.  Accordingly,
the effective potential $\Lambda_{i,j}$ becomes the matrix relative to
these quantities too.

It should be noted that a description of scattering in momentum
space is to use the quantum numbers for the relative orbital
angular momenta $\vec{l}_{i,j}$ and $\vec{\lambda}_l$ within the
pair $i,j$ and between a third body and the pair $i,j$,
respectively, together with the magnitudes of momenta
$\vec{q}_{i,j}$ and $\vec{k}_l$.

Moreover, these angular momenta have to be coupled to the total
orbital momentum $\vec{L}$ to define the partial-wave states. There
are three coupling schemes connected between each others. The
technique of determination of the recoupling coefficient in
momentum space is described in details in  \cite{Glockle}. Note
that at very low energies the most contribution has to be from S-wave
components. It simplifies the solving of problem.

The amplitudes $P_{i,j}$ have been calculated on base of equations
(\ref{Pijtot}) with different point sets, using different numerical methods
(LSARG, MATINE2 etc.). The basic features of
$n,\alpha,\alpha$-system are almost no changed. Here we give
results for case of the form-factor $\nu_{\alpha,\alpha}$ in (\ref{aSYam})
with $m = 6$.

As known in the region of positive total energy the kernels of
integral equations (\ref{Pijtot}) have "moving logarithmic singularities".
There is the method to solve this problem  \cite{Logarithm}.
However, in the present task this problem is inessential.
Upon calculation it is reasonable to take $Z = E_0 + i \Delta$ as
complex quantity with $0 < \Delta/E_0 \ll 1$.
We rely on the fact that
a little detour can not mainly misrepresent parameters of resonance
if the resonance is situated close to physical energy surface.

Calculations have been carried out in cases of very small displacements of
total energy:  $\Delta/E_0 \approx 10^{-12} \div 10^{-15}$.
It is convenient to consider the amplitude of elastic scattering
of neutron by two $\alpha$-particle resonance state $n +
(\alpha,\alpha)^{*}\rightarrow n + (\alpha,\alpha)^{*}$.

Results of
calculations confirm a fact that a set of narrow resonances appear
in low positive energy region. These resonances are situated below
the energy of $(\alpha,\alpha)^*$-resonance. Widths of these resonances
are very small.
The imaginary part of the elastic amplitude as function of
energy is shown in Fig. 1  \cite{FB19}. The values of amplitude
have been calculated in points of parameter
$x_0 = k_0/\beta_{\alpha,\alpha}$, which taken with permanent
step on $x_0$ axes: $\delta x_0/x_R \approx 0.53\cdot 10^{-3}$.

 As it is assumed positions and widths of
resonances are
almost independent on value of $\Delta$.

\begin{figure}[hbt]
  \centering
\epsfig{file=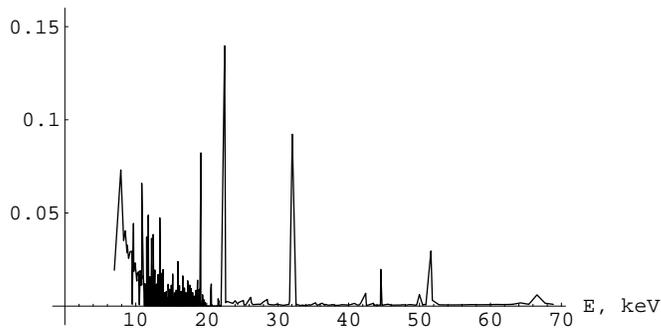,width=250pt}
\caption{Imaginary part of elastic $n, 2 \alpha$ scattering amplitude
normalized to unitary limit.}
\end{figure}

\section{The effective potential}

In order to understand the reason of resonance
phenomenon in $n,\alpha,\alpha$-system the set of equations (\ref{Pij}) can
be rewritten for elastic $n,(\alpha,\alpha)^* $ amplitude in following
form:
\begin{equation}
\label{Pii} P_{i,i}(k,k') = \Lambda_{i,i}^{eff}(k,k') + \sum_{k_s}
\Lambda_{i,i}^{eff}(k,k_s) \eta_i (Z_s)P_{i,i}(k_s,k')  \  ,
\end{equation}
where
\begin{equation}
\label{Lefii} \Lambda_{i,i}^{eff}(k,k') = \sum_{j,\ l \neq i}^3 \ \ \sum_{k_j, \
k_l} \Lambda_{i,j}(k,k_j) \widetilde{\eta}_{j,l} (Z; k_j, k_l)
\Lambda_{l,i}(k_l, k')  \ ,
\end{equation}
with
\begin{equation}
\label{Aefjl} \widetilde{\eta}_{j,l} (Z; k_j, k_l) = \eta_j (Z_j)
\delta_{j,l}  + \sum_{m \neq i} \sum_{k_m}
\eta_j (Z_j) \Lambda_{j,m}(k_j, k_m) \widetilde{\eta}_{m,l} (Z;
k_m, k_l) \ .
\end{equation}
Here $k$, $k_s$ and $k'$ are impulses of neutron in initial, intermediate
and final states, respectively.

Note that all transitions between $n,(\alpha,\alpha)^{*}$ states
marked here as $"i"$ have been collected separately in equation (\ref{Pii}).
 As
for another transitions named here as "non-elastic"
 they have been collected
apart in equations (\ref{Lefii}) and (\ref{Aefjl}). These equations define
$\Lambda_{i,i}^{eff}(k,k')$ - the effective potential of elastic
scattering $n +(\alpha,\alpha)^{*} \rightarrow
n +(\alpha,\alpha)^{*}$.

The features of $\Lambda_{i,i}^{eff}$ effective potential are mainly
formed by $\alpha,\alpha$ resonance interaction. In particular,
the ratio (\ref{kIR}) will predestine a behaviour of this potential.

1. Thus, the ratios of radii of $n,\alpha$ potentials and
$\alpha,\alpha$ resonant force prove to be very small $\xi^{J,L} =
\beta_{n,\alpha}^{J,L} / \beta_{\alpha,\alpha} \lesssim 10^{-2}$.
These ratios considered as small parameters will provoke the
decreasing of contribution of "non-elastic" transitions, i.e.
rescattering intermediate processes like $\alpha +
(n,\alpha')\rightarrow \alpha' + (n,\alpha)$. It means that we can
test the effective potential taken in the following simple form
\begin{eqnarray}
\label{LL0} \Lambda_{i,i}^{eff}(k,k') \Rightarrow
\Lambda_0^{eff}(k,k') = \{\Lambda^{eff}(\vec{k},\vec{k'})\}_{L=0} , \\
\Lambda^{eff}(\vec{k},\vec{k'}) = \sum_{\vec{k}_\alpha}
\Lambda_{i,j}(\vec{k},\vec{k}_\alpha) \eta_j(Z_\alpha)
\Lambda_{j,i}(\vec{k}_\alpha,\vec{k'})  \ , \nonumber
\end{eqnarray}
where $\vec{k}_\alpha$ is an impulse of $\alpha$-particle in intermediate
state $j\neq i$.

In other hand very small values of $\xi^{J,L}$ results in the fact
 that only the
region of very low kinetic energy turn out to be considerable in
$(n,\alpha),\alpha$ scattering channel. It means that the energy
dependent of $n,\alpha$ potentials becomes inessential. So, we can
get
\begin{equation}
\label{L0ii}
 \Lambda^{eff}(\vec{k},\vec{k'})\Rightarrow \overline{\Lambda}(\vec{k},\vec{k'})
= \sum_{\vec{k}_\alpha} <\nu_{\alpha,\alpha}| G_0 (Z) \ C_{n,\alpha} \ G_0 (Z)
|\nu_{\alpha,\alpha}>   \ ,
\end{equation}
where $C_{n,\alpha} =  <y|\nu_{n,\alpha'}> \eta_{n,\alpha'}(Z_\alpha)
<\nu_{n,\alpha'}|y'> $ \ is closely a constant because
$y = x\cdot \xi^{J,L} \ll x$ and $y' = x' \cdot \xi^{J,L} \ll x'$.

Therefore, we find easily $\overline{\Lambda}$ \ in the analitical form.

For example, for $S$-wave we come to the expression ($x = k/\beta_{\alpha,\alpha}$,
$x' = k'/\beta_{\alpha,\alpha}$, see above)
\begin{equation}
\label{L01}
 \overline{\Lambda}_{L=0}(x,x') = \frac{Const}{x\cdot x'} \widetilde{\Lambda}
\ , \  \  \ \ \
\widetilde{\Lambda} = \frac{F_1+F_2+F_3+F_4}
{(1+Z_i)^3\cdot(1+Z'_i)^3}   \ ,
\end{equation}
where $Z_i = Z/\beta^2_{\alpha,\alpha}-9(x)^2/4$, \
$Z'_i = Z/\beta^2_{\alpha,\alpha}-9(x')^2/4$, \ and
$Const = N_{\alpha,\alpha}^2 N_{n,\alpha}^2/2\pi^2$.

Introducing $d_r = (x-x')/2$ and $d_m = (x+x')/2$ we can get
\begin{equation}
\label{F1}
F_1 = \frac{1}{2} \sum_{n=1}^{4}(-1)^n R(a_n) \ ,
\end{equation}
where
\begin{equation}
\label{Ra}
R (a_n) =
\frac{a_n}{2} lg \frac{a_n^2+d_r^2}{a_n^2 + d_m^2}
+d_m\cdot tg^{-1}(d_m/a_n)- d_r\cdot tg^{-1}(d_r/a_n)  \ ,
\end{equation}
with  $a_1 = \kappa + 1$, $a_2 = 1 + 1$, $a_3 = 1 + \kappa'$ and
$a_4 = \kappa + \kappa'$, where $\kappa = \sqrt{-Z_i}$,
$\kappa' = \sqrt{-Z'_i}$.

In case of $Z_i > 0$ or $Z'_i > 0$ we must
substitute $\kappa\rightarrow -i\cdot\sqrt{Z_i}$ and
$\kappa' \rightarrow -i\cdot\sqrt{Z'_i}$.

The function $F_2$ can be written in the following form
\begin{equation}
\label{F2}
F_2 = \frac{1}{4}(M_1 + M_2) \ ,
\end{equation}
where
\begin{equation}
\label{M1}
M_1 = [lg \frac{a_1^2+d_r^2}{a_1^2 + d_m^2} -
lg \frac{a_2^2+d_r^2}{a_2^2 + d_m^2}]\cdot[(1+Z'_i)/2 + (1+Z'_i)^2/8] \ ,
\end{equation}
and
\begin{equation}
\label{M2}
M_2 = [lg \frac{a_3^2+d_r^2}{a_3^2 + d_m^2} -
lg \frac{a_2^2+d_r^2}{a_2^2 + d_m^2}]\cdot[(1+Z_i)/2 + (1+Z_i)^2/8] \ .
\end{equation}
$F_3$ can be written as
\begin{equation}
\label{F3}
F_3 = - K_1 + K_2 - K_3 \ ,
\end{equation}
with
\begin{eqnarray}
\label{K1}
K_1 & = & \frac{a_1}{2}\cdot[\frac{1}{a_1^2+d_r^2}-\frac{1}{a_1^2+d_m^2}]
\cdot\frac{(1+Z'_i)^2}{8} \ , \\
\label{K2}
K_3 & = & \frac{a_3}{2}\cdot[\frac{1}{a_3^2+d_r^2}-\frac{1}{a_3^2+d_m^2}]
\cdot\frac{(1+Z_i)^2}{8} \ , \\
K_2 & = & (K_{2,0} + K_{2,1})\cdot [\frac{1}{a_2^2+d_r^2}-\frac{1}{a_2^2+d_m^2}],
\end{eqnarray}
where
\begin{eqnarray}
\label{K21}
K_{2,0} & = & (\frac{(1+Z_i)^2}{8}+\frac{1+Z_i}{2})\cdot
(\frac{(1+Z'_i)^2}{8}+\frac{1+Z'_i}{2}) \ , \\
K_{2,1} & = & \frac{(1+Z_i)^2}{8}\frac{(1+Z'_i)^2}{8} \ ,
\end{eqnarray}
and
\begin{equation}
\label{F4}
F_4 = B\cdot[\frac{a_2^2-d_r^2}{(a_2^2+d_r^2)^2}-
\frac{a_2^2-d_m^2}{(a_2^2+d_m^2)^2}]
+H\cdot[\frac{1-\frac{3}{4}d_r^2}{(a_2^2+d_r^2)^3}-
\frac{1-\frac{3}{4}d_m^2}{(a_2^2+d_m^2)^3}] ,
\end{equation}
with
\begin{eqnarray}
\label{F4B}
B & = & \frac{1}{2}(B_i + B'_i) \ , \\
B_i & = & (\frac{(1+Z_i)^2}{8}+\frac{1+Z_i}{2})\cdot(\frac{(1+Z'_i)^2}{8}) \ , \\
B'_i & = & (\frac{(1+Z_i)^2}{8})\cdot(\frac{(1+Z'_i)^2}{8}+\frac{1+Z'_i}{2}) \ , \\
H & = & \frac{1}{8}(1+Z_i)^2\cdot(1+Z'_i)^2 .
\end{eqnarray}

\vspace{0.5cm}

2. It's easy to see that $\widetilde{\Lambda}$ remains a finite value
 in limit \
$Z_i \rightarrow - 1$ \ and (or) \ $Z'_j \rightarrow - 1$.
It happens due to the fact that the sum of $F_n$ in numerator of (\ref{L01})
is coming to zero in this limit too.

In limit $x \rightarrow 0$ and (or) $x' \rightarrow 0$ the quantity
$\widetilde{\Lambda} \rightarrow 0$ as well.
If $Z_i$ \ and (or) \ $Z'_j$ are increasing to big values the quantity
$\widetilde{\Lambda}$ will be quickly decreasing.

Moreover, the effective potential - $\widetilde{\Lambda} (x, x')$
is vanishing almost everywhere
excluding narrow strips close to values
$x, x' = 2/3 \sqrt{1 + E_0/\beta^2_{\alpha,\alpha}}$.
 Within these areas the effective potential can
achieve large values. Figure 2 shows the behaviour of
$\widetilde{\Lambda} (x, x')$.

\vspace{0.5cm}

Note, that first point of conclusion leads to the fact that
$n,\alpha$ interaction turns out as the contact one. And the second point
signifies that $\alpha,\alpha$-system acts like the resonator.
Similar effects are well-known in quantum physics.

\begin{figure}[hbt]
\centering
\epsfig{file=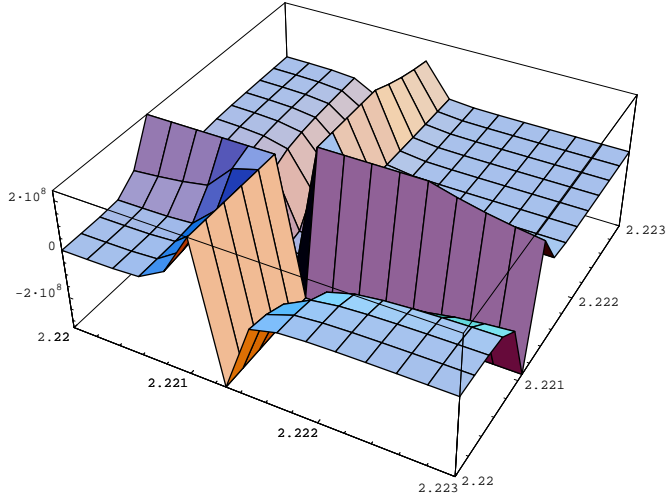,width=250pt}
 \caption{The real part of effective potential
$\widetilde{\Lambda}(x,x')$ at  $Z = 0.9 keV$}
\end{figure}

It is easy to see that the $n,(\alpha,\alpha)$ effective potential
comes to the form close to well-known pseudopotential approach \cite{Baz},
 \cite{Demkov}.
For example, the task on scattering of particle by system of two
heavy centers can be recalled. It is interesting the case of
 repulsive interactions between
the particle and every of both centers with a $\delta$-function
form in coordinate space.

It is obvious that two-body subsystem of
particle and one heavy center can not form any bound state.
However, even in the case the three body system consisting of the
particle and two heavy centers will have a bound state  \cite{Baz}.

Here we consider a resonator instead of heavy centers system, the
simplest resonator being a resonant state of two $\alpha$-particles.
And we consider "the neutron -
neutral particle" as a scattering particle.
This fact represents a very important point.

It is remarkable that in every of above mentioned cases the scattering
particle will be ensnared by subsystem. In case of neutron
scattered by the resonator it can be said
 that the neutron will be in state of total inner reflection between
$\alpha$ particles. Lifetime of this three body state will be
limited by
 two $\alpha$-particles subsystem own lifetime.

 The resonance state of three $\alpha$-particles
 can be assumed in this model as more complex resonator.

It should be noted that pseudopotential of scatterring particle by
system of few heavy centers leads to the bound states too, i.e. a
discrete spectrum can exist in similar systems  \cite{Baz}. Thus,
it is additional argument to suggest that the
system consisted of few $\alpha$-particles and one neutron has a very
reach spectrum.

Calculations have been carried out and give the following results.

In case of $E > 0$ calculations confirm an existence
of narrow resonance set at very low energies.
The amplitude $\overline{P}_{L=0; i,i}$ from (\ref{Pii}),
with $ \Lambda^{eff}_{L=0; i,i} = \overline{\Lambda}_{L=0}(x,x')$,
have been determined for point set of $x_0$ as above.

Figure 3 shows the curve of real part of $\overline{P}_{L=0; i,i}$
 in energy region $1.5 \div 3 \ keV$.

\begin{figure}[h]
  \centering
\epsfig{file=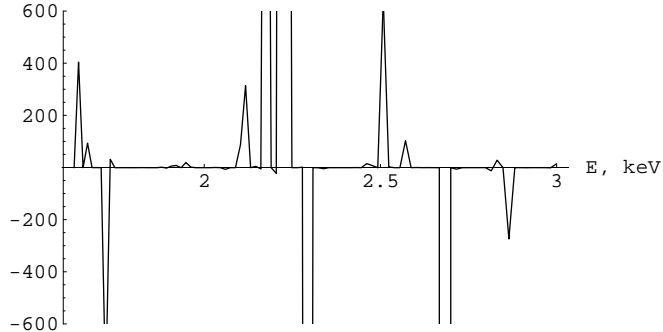,width=250pt}
 \caption{The real parts $\overline{P}_{0; i,i}(x,x')$
 in region $Z > 0$, $x = x' = 0.06$}
\end{figure}

In case of $Z < 0$, i.e. in
the region of discrete spectrum, $n(\alpha,\alpha)^{*}$-system
has the bound state with very small energy $E_B \simeq - 4.62 \ keV$.

It is very interesting fact that in accordance with results of well-known
problem for a particle scattering by two heavy centers (see, above).
It is clear that lifetime of this bound state is limited by the own
lifetime of $\alpha,\alpha$-resonance.

\section{Remarks on astrophysical aspects}

It seems there are two serious things which can prevent
the resonance fusion: Coulomb forces and nuclear reactions of
neutron capture by few $\alpha$-particles subsystem.

Our calculations carried out with Coulomb force between two
$\alpha$-particles have confirmed the fact of resonance
phenomenon in this case just as the resonance
level shifts.

We are going to continue calculations using more precise numerical methods.

As for nuclear reactions of neutron capture a situation is not
clear in few body dynamics at very low energies. First of all it is
interesting to study a neutron capture directly by few
$\alpha$-particle subsystem. We know that one $\alpha$-particle can
not capture a neutron.

Introduce $N_f$ - average of fusion acts stimulated by neutron in
$\alpha$-particle matter. We assume that main part of time between
two fusion acts has to be taken by neutron slowing in
matter.

Remind that a lifetime of free neutron  $\sim 10^3$ s, which is
infinity in comparison with average time of nuclear fusion, average time
of neutron slowing, or lifetime of $\alpha,\alpha$-resonance.
Therefore, we can give a rough estimation for
$N_f$ as ratio between probabilities of $\alpha$-particles fusion and
neutron capture: $N_f = W_\alpha/W_n$.

Following a qualitative reasoning we can quess that probability of
neutron capture by system of few $\alpha$-particles has to be more
less than probability of direct fusion in few $\alpha$-particles
subsystem. In last case the neutron becomes free:
this is a result from a considerable profit of energy.
Moreover, when happened a fusion the few-body system is easy to throw
neutron out as the lightest particle than more heavy $\alpha$-particles.

In any case we can talk over the resonance fusion phenomenon.
If resonant reactions are effective in stellar matter they present
us a possibility to explain a periodic activity of stars, in particular
periodic local explosions at the Sun.

As for a problem of free neutrons in stellar matter we can remind
about reactions with lightest nuclei, like $d + d \rightarrow ^3He + n + Q$
and $d + T \rightarrow \alpha + n + Q$, etc. Thus, an existence of free
neutrons inside the Sun is possible.

Then, assuming that the small part of the Sun energy is a result of resonance
fusion we can understand the problem of little neutrino deficiency incoming to
the Earth from the Sun.

It is clear that a lot of interesting effects and phenomena arise from
the resonance fusion. Note that the resonance fusion can be easily
put into practice to generate energy.

Next article will concern new application projects of resonance fusion.

\section{conclusions}

In conclusion we summerise our results.

The $n,\alpha,\alpha$-system has a set of narrow resonances in low
energy region. Their lifetime is comparable to ground
$\alpha,\alpha$-resonance.

Interacting with a free neutron the $\alpha,\alpha$-system can act
like the resonator.
And the $n,(\alpha,\alpha)^*$ effective potential
has the form close to well-known pseudopotential which describes
the scattering of particle
 by repulsive $\delta$-function forces of two
heavy centers. As it is known this pseudopotential gives
a bound state of this three-body system.

So, we can say about the new type of fusion -
resonance fusion. The neutron can play a role of catalyst
to stimulate $\alpha$-particles fusion.

The number of fusion acts stimulated by one
neutron may reach large values because of neutron huge lifetime
$\sim 10^3 \ s$ in comparison with nuclear interaction times. By
realizing possibility of light nuclei generation
in resonance fusion,
we can give new answers to questions of origin
 of light nuclei.

The resonant behavior of reactions may explain
periodic activity of stars and periodic local
explosions at the Sun. Since it means resonant dependence of
energy production on local
temperature inside the Sun.

Processes
of resonance fusion can be investigated experimentally even
nowadays. One of existing and operating installations could be used
for.

\end{document}